# Second Order Phase Transition and Universality of Self-Buckled Elastic Slender Columns

Desyana Olenka Margaretta, Nadya Amalia, Fisca Dian Utami, Sparisoma Viridi, and Mikrajuddin Abdullah[a]

Department of Physics
Bandung Institute of Technology
Jalan Ganesa 10 Bandung 40132, Indonesia
[a]Email: mikrajuddin@gmail.com

**Abstract**

Self-buckling is an interesting phenomenon that is easily found around us, either in nature or in objects made by human. Palm fronds which initially directed upward when they were short and turned into bending after appreciably longer is an example of the self-buckling phenomenon. We report here that the self-buckling of columns can be treated as a process of second order phase transition by considering the straight column as "disorder state", the bending column as "order state", and the temperature as the inverse of column length. The "critical temperature" corresponds to the inverse of critical length for buckling, $1/L_{cr}$, and the deviation angle made by column free end relative to vertical direction satisfies a scaling relationship with a scaling power of 0.485. Changing of the column geometry from the vertically upward to the bending state can be considered as a transition from disorder state to order state.

# 1. Introduction

Initially, most *paddy* leaves grow up vertically, but after a certain height, they start to bend. We observe the similar phenomenon in the *pandanus* leaves or other leaves having similar shape (long and slim). The similar phenomenon is also observed in hairs growing from bald or nearly bald heads or skins. The hairs, initially grow straight and then bend after reaching a certain length.

The *banana*'s young leaf initially grows up and stays vertically in a shape like a cylinder. But, when the leaf sheet opens, the leaf suddenly bends. The palm's young leaf also shows the same mechanism, initially grows vertically upward and then bends when the leaf opens. Those are all very common phenomena around us, but most of us have likely ignored them and considered neither interesting scientific role involved. At this work we will show that the physical roles underlying such phenomena are very attractive. Surprisingly, such phenomena mimic a kind of phase transition.

To demonstrate such phenomena more viable, we can use a sheet clamped between two vertical plates. The sheet is then moved up progressively through the clamp. We clearly observe the sheet initially directs vertically. But, by increasing the height of the sheet above the clamp, the sheet suddenly bends having reached a certain height, after which the sheet continuously bends (**Figure 1(a))**. This process is similar to bending of *paddy* or *pandanus* leaves when reaching a certain length. **In Fig. 1(b)** we show the variation of bending profiles of the paper sheets having different lengths fixed together at a vertical frame. Simetrical bending formation was clearly observed.

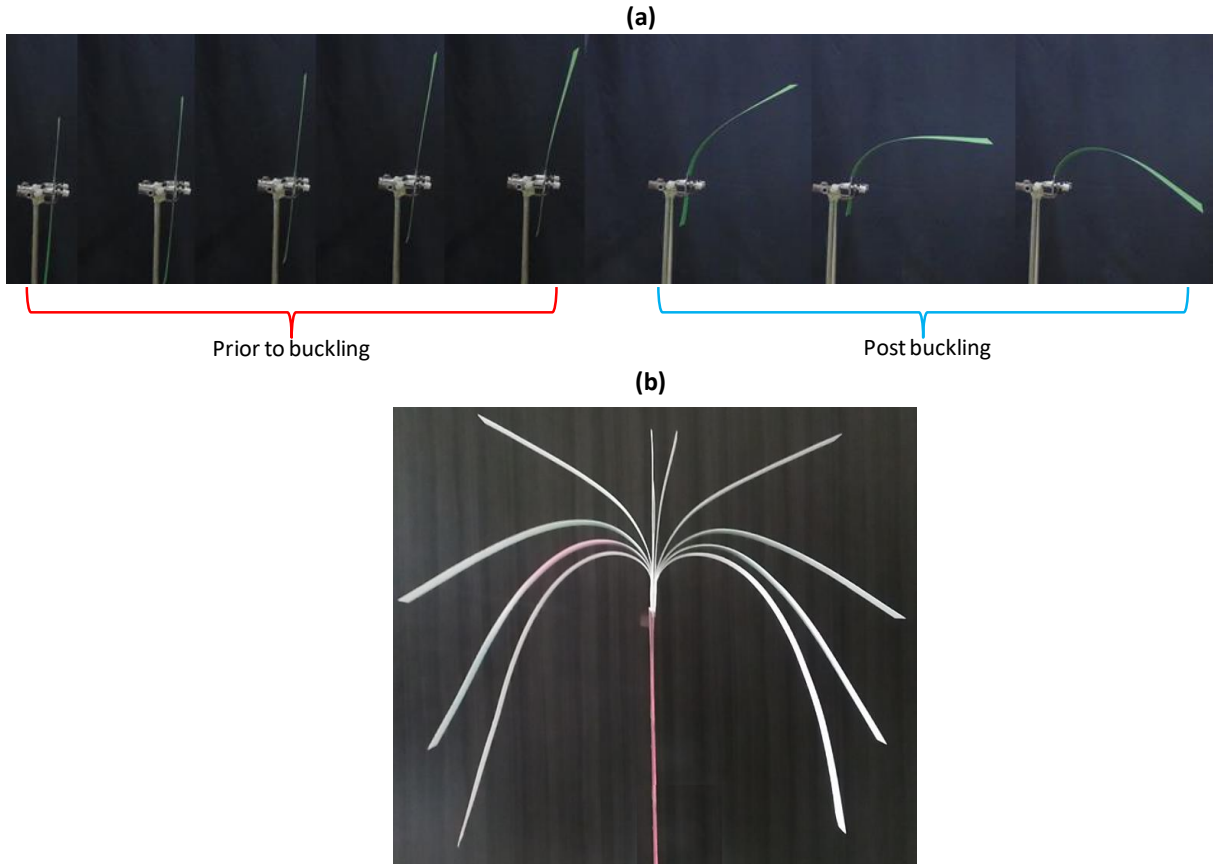

**Figure 1** (a) Bending evolution a paper that progressively pushed up through a vertical clamp: (left) prior to buckling and (right) after buckling. (b) bending variation of sheets having different lengths fixed together at a vertical frame.

*Paddy* leaves, *pandanus* leaves, *banana* leaves, and sheet are examples of slender columns. Bending of a slender column is a process of self-buckling. It is then interesting to reconstruct the bending evolution of the sheet after experiencing a self-buckling and this attempt is rarely discussed by authors.

In fact, many common phenomena around us have challenged researchers. After the phenomena are examined in-depth, many unexpected behaviors are identified. It could happen from such explorations, unexpected new technologies or new tools were born. For example, understanding the mechanism of formation of ring stains from dried liquid drops open the opportunity to develop high-quality printer inks [1]. Reis group from MIT has published many papers describing likely common phenomena around us and some of them have technological impacts [2-8].

Most discussions regarding the self-buckling were merely focused on the boundary problems. At the present work, we extract many attractive phenomena in the self-buckling process that likely rarely considered. For this purpose, we use the discrete form of the bending equation of slender columns, rather than the continuum form, and the calculation can be easily processed numerically. The discrete equation can be applied for describing bending profiles of slender columns having arbitrary bending angles (either small or large bending). This approach might be compared to the direct approach, which is based on the deformable curve model as reported by Bîrsan et al [9,10]. This is an efficient approach for analyzing elastic beams and rods with a complex internal structure (functionally graded, composite, nonhomogeneous, etc.). Linul et al [11-13] and Berto et al [14,15] have applied this method to describe bending of various composites (inhomogeneous) beams. This method is also important for describing the bending of materials where the mechanical properties can be controlled. Study of the change in the mechanical behavior of materials when exposed to radiation (neutron, ion, or electron beams) was reported by Zeyad [16].

In the present work, however, we were able to relate the shape of the self-buckled sheet with a phase transition phenomenon. In addition, we also identified a universal equation for describing the bending for all homogeneous columns and proved the famous critical column height for buckling, $L_{cr} = (7.8373 YI/\rho gA)^{1/3}$ ($Y$ = elastic modulus, $I$ = area moment, $\rho$ = density, $A$ = cross-section area, and $g$ = the acceleration of gravitation) [17], in a strongly different approach. This is not a hard topic and might be not a very new one. It is likely a matter of critical thinking, observing phenomena around and imagining the physical roles underlying them. Recently, Rotter et al reported investigation of nonlinear stability of thin elastic cylinders of different length under global bending **[18,19]** which is likely further deep investigation of the buckling phenomenon.

## 2. Method

At present we restrict our consideration to inextensible columns without twisting. This assumption is very acceptable for geometry like a sheet. The equilibrium equation is $\vec{M} = -(YI/R)\hat{t} \times \hat{n}$, with $\vec{M}$ flexion moment, $R$ the curvature which can be expressed as $1/R = |d\hat{t}/ds|$ or $1/R = d\theta/ds$, $\hat{t}$ the unit vector along the centerline, and $\hat{n}$ the unit normal vector perpendicular to $\hat{t}$. We choose $\hat{t}$ and $\hat{n}$ to govern the $x$-$y$ plane ($x$ axis to right and $y$ axis to vertical upward). The $\vec{M}$ directs parallel to the $z$ axis and after performing a scalar product with $\hat{e}_z$, the bending equation reads

$$M_z = -\frac{YI}{R} \tag{1}$$

*2.1 Discretization*

We need to transform Eq. (1) into a discrete form so that it can be easily processed numerically. **Figure 2** illustrates a column having the chord length $L$, divided into $N$ identical segments, $a$ (= $L/N$). The segment at the free end is the 1st segment, and at the clamp (fixed end) is the $N$-th segment. The $j$-th segment makes an angle $\theta_j$ ($j$ =1 to $N$) to the horizontal.

In general, the flexion moment at the $j$-th segment is contributed by weights of all segments and non-gravitational to the left. We assume the column bends counter clockwise when moving from the fixed end to the free end. The position of the $i$-th segment relative to the $j$-th segment ($i < j$) is $\vec{R}_{ij} = \hat{e}_x x_{ij} + \hat{e}_y y_{ij}$, where $x_{ij} = -a\sum_{k=i}^{j-1}\cos\theta_k$ and $y_{ij} = -a\sum_{k=i}^{j-1}\sin\theta_k$. The $i$-th segment experiences a gravitational force of $\vec{W}_i = -\lambda_i a g \hat{e}_y$ and arbitrary non-gravitational force $\vec{F}_i$, both produce in the flexion moment at the $j$-th segment, $\vec{M}_{ij} = \vec{R}_{ij} \times (\vec{F}_i + \vec{W}_i)$. The total flexion moment at the $j$-th segment from all segments to the left ($i$=1 to $j$-1) becomes

$$\vec{M}_j = \sum_{i=1}^{j-1}\vec{R}_{ij} \times \vec{F}_i + a^2 g \hat{e}_z \sum_{i=1}^{j-1}\sum_{k=1}^{i}\lambda_k \cos\theta_i \tag{2}$$

The component of this moment to the z direction is $M_{jz} = \hat{e}_z \bullet \vec{M}$.

Eq. (1) can be discretisized as $M_{jz} = -Y_j I_j (\theta_j - \theta_{j-1})/a$, with $I_j$ and $Y_j$ are the area moment and the young modulus of the $j$-th segment, respectively (generally depending on position), and the bending equation becomes

$$\theta_j = \theta_{j-1} - \frac{a}{Y_j I_j}\sum_{i=1}^{j-1}\hat{e}_z \bullet (\vec{R}_{ij} \times \vec{F}_i) - \frac{a^3 g}{Y_j I_j}\sum_{i=1}^{k-1}\sum_{k=1}^{i}\lambda_k \cos\theta_i \tag{3}$$

*This equation is a general equation for arbitrary columns, either homogeneous or inhomogeneous.*

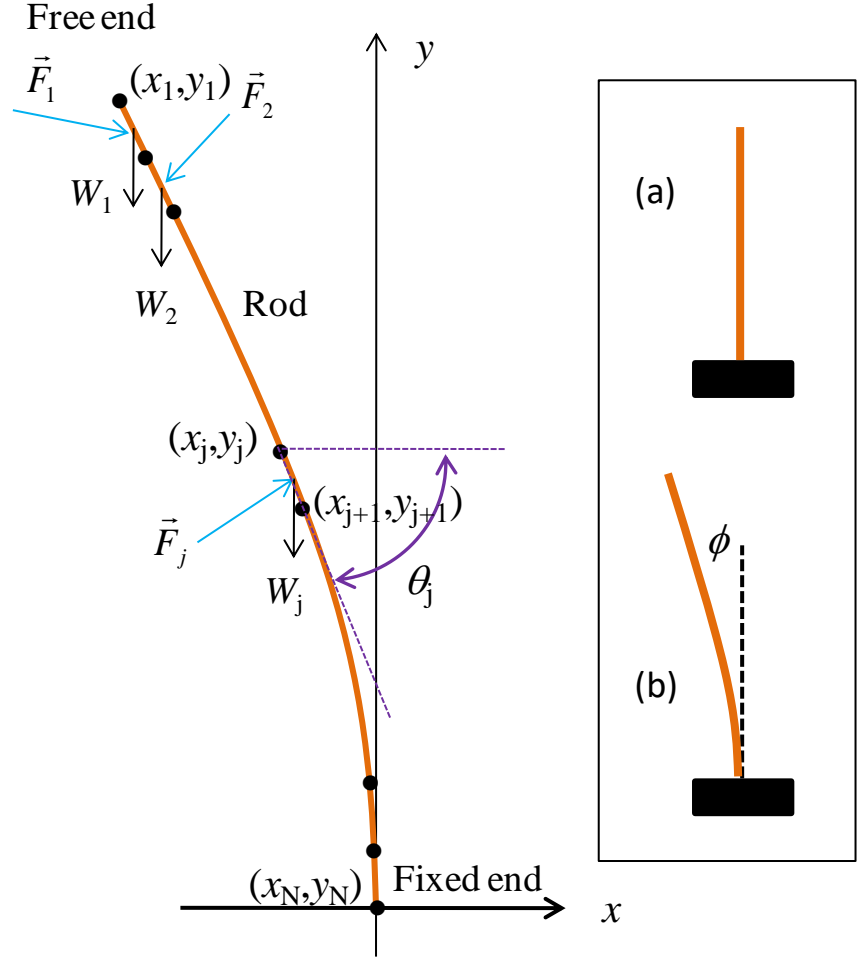

**Figure 2.** (left) Schematic of a column divided into *N* identical segments and segment numbering. (right) Conditions of column: (a) before buckling and (b) right after buckling.

For a special case when the external force is a point force acting to the *n*-th segment only, $\vec{F}_i = (F_{nx}\hat{e}_x + F_{ny}\hat{e}_y)\delta_{in}$, with $\delta_{in}$ is the Kronecker delta (equal to 1 if $i = n$ and equal to 0 if $i \neq n$) and the last bending equation reduced to

$$\theta_j = \theta_{j-1} - \frac{a^2}{Y_j I_j}\sum_{k=n}^{j-1}(F_{nx}\sin\theta_k - F_{ny}\cos\theta_k) - \frac{a^3 g}{Y_j I_j}\sum_{i=1}^{j-1}\sum_{k=1}^{i}\lambda_k \cos\theta_i \qquad (4)$$

When the external force is a certain load suspended to the *n*-th segment, $F_{nx} = 0$ and $F_{ny} = -W_n$, and the equation becomes

$$\theta_j = \theta_{j-1} - \frac{a^2 W_n}{Y_j I_j}\sum_{k=n}^{j-1}\cos\theta_k - \frac{a^3 g}{Y_j I_j}\sum_{i=1}^{j-1}\sum_{k=1}^{i}\lambda_k \cos\theta_i \qquad (5)$$

When the force acting on the rod is only its own weight, $\vec{F}_i = 0$ for all $i$, we obtain a simple relation

$$\theta_j = \theta_{j-1} - \frac{a^3 g}{Y_j I_j} \sum_{i=1}^{j-1} \left( \sum_{k=1}^{i} \lambda_k \right) \cos \theta_i \quad (6)$$

Equation (6) has been used by us to determine the elastic modulus of slender beams by processing the bending image of the beam. This is possible since, although the equation has been derived for the column, it also applies for beams. The method is also potential for estimating the glass transition temperature of polymeric materials [20]. Very precise glass transition temperatures of several polymers have been obtained.

## 2.2 Numerical Procedures

Let us first restrict on the case when the column is bent by its own weight. We will calculate angles of all segments by firstly fixing the angle for the first segment. The problem is, the boundary condition is applied to the fixed segment (the last segment), instead of the 1st segment. Therefore, eventually, after performing calculations the angle made by all segments, the calculated angle of the fixed segment is no longer satisfies the boundary condition. If this result happens, the initial angle applied to the first segment was wrong and we must try another angle for the first segment until the calculated angle of the last segment is equal to the boundary condition. This approach looks tedious, but we can simplify the solving process by using iteration. The criterion for stopping the iteration is $|\theta_N - \theta_{bc}| < \varepsilon$, with $\theta_{bc}$ is the bending angle of the *N*-th (fixed segment) and $\varepsilon$ is a small number which depends on the accuracy we are intending. In simulation we used $\varepsilon = 0.001$ rad.

**Figure 3** illustrates the flow chart for calculating the true bending angles of all segments. After obtaining the true set of angles { $\theta_1, \theta_2, \theta_3, \theta_4, ..., \theta_N$ }, we then determine the coordinates of each segment calculated using $x_j = x_{j-1} + a \cos \theta_j$ and $y_j = y_{j-1} + a \sin \theta_j$.

*2.3 Experiments*

We have inspected the evolution of the sheet shape as the function of its height. The first attempt used an A4-sized photocopy paper ($\ell$ = 0.297 m length, $w$ = 0.21 m width, and $\delta$ = $10^{-4}$ m thickness) of mass density per unit area $\sigma$ = 70 gsm (0.07 kg/m$^2$) or the mass density per unit length $\lambda = \sigma w$ = 0.0147 kg/m. The area moment of the paper is $I = w\delta^3/12$. The reported elastic modulus of paper is around $Y$ = 3 GPa [21,22]. The predicted critical height of this paper derived from the Euler-Bernoulli equation is 0.143 m.

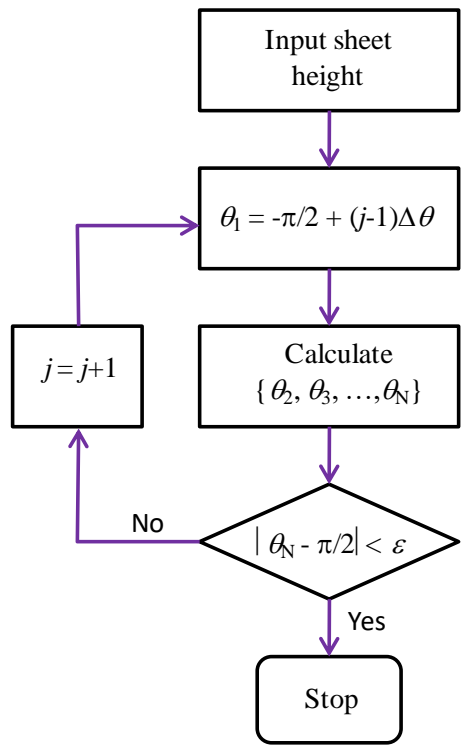

**Figure 3.** Flow chart for calculating the true bending angles of all segments.

## 3. Results and Discussion

*3.1 Phase Transition*

**Figure 4(a)** shows the calculated paper shape at different heights using Eq. (6) and data for 70 gsm photocopy paper. We calculated for papers of different lengths: (1) $L$ = 0.22 m, (2) $L$ = 0.19 m, (3) $L$ = 0.17 m, (4) $L$ = 0.157 m, (5) $L$ = 0.15 m, (6) $L$ = 0.144 m, (7) $L$ = 0.142 m, (8) $L$ = 0.14185 m, and (9) $L$ < 0.1418 m. We observed the paper largely bends when the length is very high (1). The bending decreases when the length is reduced. Bending with $\theta_l$ positive occurs when $L > 0.157$ m (states 1 to 4), and bending with $\theta_l$ when occurs when $0.14.1$ m $< L < 0.157$ m (states 4 to 9). By carefully varying the paper height, we observed that the critical height where the paper suddenly directs straight upward. The starting height for shaping straight vertical is exactly the same as the critical buckling height 0.1418 m. The $\theta_l$ is unchanged when $L$ just is shorter than $L_{cr}$.

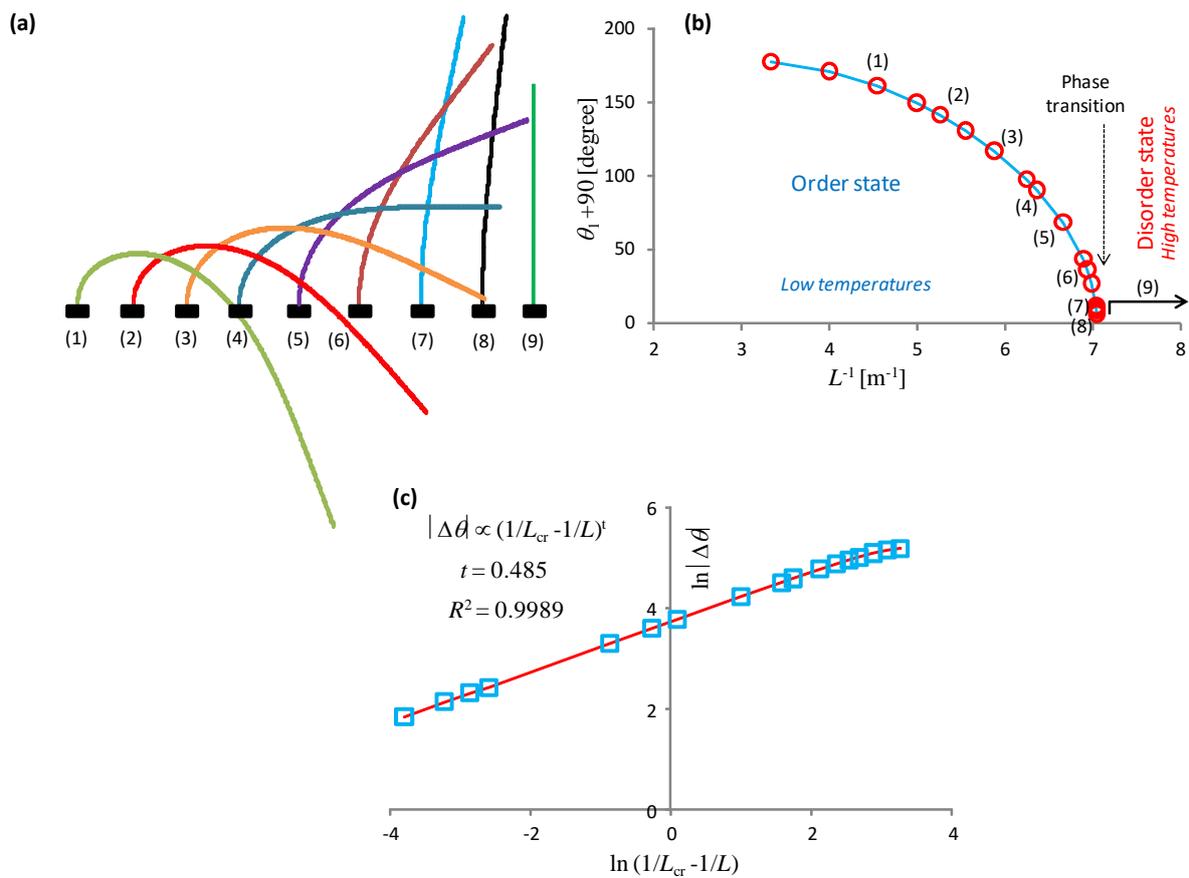

**Figure 4**. (**a**) Bending profile of 70 gsm photocopy paper directed vertically at different lengths: (1) $L = 0.22$ m, (2) $L = 0.19$ m, (3) $L = 0.17$ m, (4) $L = 0.157$ m, (5) $L = 0.15$ m, (6) $L = 0.144$ m, (7) $L = 0.142$ m, (8) $L = 0.14185$ m, and (9) $L < 0.1418$ m. (**b**) The angle made by the free end (added with 90º) as the function of paper height. See text for description of $\theta_l$. (**c**) Plot of $\ln|\Delta\theta_l|$ with respect to $\ln(1/L_{cr} - 1/L)$. Symbols are simulation results and the line is a fitting curve.

**Figure 4(b)** shows the dependence of $\theta_1+90°$ on the paper height that produces $\theta_N = -\pi/2$ (-90°). Only one $\theta_1$ that produces $\theta_N = -90°$ for a certain paper height. This value corresponds to equilibrium shape of the paper. Curve was plotted with respect to $1/L$ (correspondence to temperature in phase transition). In phase transition, the order state of zero occurs at high temperature, while in our case, the free end angle of zero occurs at high $1/L$ (short $L$). The change in **Fig. 4(b)** is very similar to second order phase transition, where the first phase (order) corresponds to bending state ($1/L < 1/L_{cr}$) and the second phase (disorder) corresponds to upward straight shape ($1/L > 1/L_{cr}$). The phase transition occurs at $1/L = 1/L_{cr}$ (corresponds to the critical temperature in phase transition).

To further inspect the phase transition behavior, we plot $\ln|\Delta\theta_1|$ as the function of $\ln(1/L_{cr} - 1/L)$ where $|\Delta\theta_1|$ is the deviation of angle pointed by the free end relative to the angle vertical position as shown in **Fig. 4(c).** We obtain a truly straight line, strong showing a scaling relation

$$|\Delta\theta_1| \propto (1/L_{cr} - 1/L)^t \tag{7}$$

with the critical exponential $t = 0.485$ (could be rounded to 0.5). We will simply prove the evidence of this scaling relationship as follows.

By carefully inspecting **Fig. 4(b),** it is clear that the curve is very similar to curve of order of parameter in second order phase transition where $1/L$ plays a role as temperature and $\Delta\theta_1+90$ (in degree) is the order parameter. Therefore, using the Landau theory for the second order phase transition we may approximate the bending column free energy as [23]

$$f(1/L) = f_0(1/L) + \alpha m^2 + \frac{1}{2}\beta m^4 \tag{8}$$

with $m = \Delta\theta_1 + 90$ (in degree), $\alpha$ and $\beta$ are positive constants, and $f_0$ is the free energy at very high $1/L$ (very short column), corresponds to very high temperature.

Similarly, taking the first order in "temperature" (inverse of column length) expansion of $\alpha \approx \alpha_0(1/L - 1/L_{cr})$ and keeping the parameter $\beta$ to nearly constant, we have

$$f(1/L) \approx f_0(1/L) + \alpha_0(1/L - 1/L_{cr})m^2 + \frac{1}{2}\beta m^4 \tag{9}$$

The order parameter that minimizing the free energy satisfies

$$df/dm = 0, \quad (10)$$

resulting,

$$m = \begin{cases} 0 & if \quad 1/L > 1/L_{cr} \\ \propto (1/L_{cr} - 1/L)^{1/2} & if \quad 1/L < 1/L_{cr} \end{cases} \quad (11)$$

which is similar to the expression in Eq. (7).

The zero temperature state corresponds to $1/L \to 0$ or $L \to \infty$. In this case, the column free end directs vertically downward. This state can be considered as the most ordered state. When temperature is increased or $L$ is reduced, the angle of the free end decreases and becomes directly upward when $1/L \to 1/L_{cr}$, and then stays zero when $1/L > 1/L_{cr}$.

## 3.2 Universality for Self-Buckling Column

We will show that Eq. (2) is universal, applied for all homogeneous slender columns having constant area moment mass density. For such columns, we have $I_j = I$, $Y_j = Y$, and $\lambda_j = \lambda$ for all $j$. Therefore, Eq. (2) can be re-expressed as

$$\theta_j = \theta_{j-1} - \frac{a^3 \lambda g}{YI} \sum_{i=1}^{j-1} i \cos \theta_i. \quad (10)$$

For very large number of segments, or very small $a$, the result of numerical calculation does not change even if we use different segment size, either larger or smaller than the previous one. This conclusion is simply demonstrated in **Appendix 1**.

Now let us consider another column made of different material having $Y'$, $I'$, and $\lambda'$. Since we can arbitrarily select $a$ as long as still produce a large number of segments, let us select $a'$ so that $a^3 \lambda g / YI = a'^3 \lambda' g / Y'I' = \xi$ (= constant) for all materials. Therefore, the segment length must satisfy

$$a = \left(\frac{\xi YI}{\lambda g}\right)^{1/3}. \tag{11}$$

Based on this selection, the recursive Eq. (10) can be re-expressed as

$$\theta_j = \theta_{j-1} - \xi \sum_{i=1}^{j-1} i \cos \theta_i. \tag{12}$$

Equation (12) is independent of material, since $\xi$ is constant, therefore it is a universal equation.

The angle made by the fixed end is $\theta_N = \theta_{N-1} - \xi \sum_{i=1}^{N-1} i \cos \theta_i$. This angle depends on $\theta_1$ as the starting point of recursion. Since this equation is universal, the condition for the first segment angle to start to deviate from the vertical position will occur at the same $N$ (the same segment number) for all columns. Thus, for all materials, the buckling occurs at the same $N_{cr}$, satisfying $(-\pi/2) = \theta_{N_{cr}-1} - \xi \sum_{i=1}^{N_{cr}-1} i \cos \theta_i$. But, even when different materials have the same $N_{cr}$, they actually have different length since the lengths of segments are different. Thus material with $Y$, $I$, $\lambda$ and material with $Y'$, $I'$, and $\lambda'$ have the same $N_{cr}$. We then have $L_{cr} = N_{cr} a$ and using Eq. (11) we can then write $L_{cr} = N_{cr}(\xi YI / \lambda g)^{1/3}$. Since, $\lambda = \rho A$, we finally have

$$L_{cr} = \left(N_{cr}^3 \xi \frac{YI}{\rho A g}\right)^{1/3} \tag{13}$$

It must be noted that $N_{cr}$ is identical for all materials (the solution of universal Eq. (12) that making $\theta_N = -\pi/2$) hence it is a constant. Eq. (15) is precisely equal to the critical column height for self-buckling that has been derived two centuries ago [17] by selecting $N_{cr}^3 \xi = 7.8373$.

Now, let us define a characteristic length as

$$L_0 = \left(\frac{YI}{\lambda g}\right)^{1/3} \tag{14}$$

For example, the corresponding $L_0$ for various sheets are shown in **Table 1**. It is easy to show that $L_{cr} = 2L_0$. We express the column length in such length scale as $\tilde{L} = L/L_0$. By remembering the definition of $a = L/N$ we can rewrite Eq. (12) as

$$\theta_j = \theta_{j-1} - \left(\frac{\tilde{L}}{N}\right)^3 \sum_{i=1}^{j-1} i \cos\theta_i. \tag{15}$$

**Table 1**. Mechanical parameters of samples.

| Materials | Thickness (m) | Width (m) | Linear density (kg/m) | Area moment (× $10^{-14}$ m$^4$) | Elastic modulus (Gpa) | $L_0$ (m) |
|---|---|---|---|---|---|---|
| Copy paper 70 gsm | 0.0001 | 0.21 | 0.0147 | 1.75 | 3[a] | 0.071 |
| Copy paper 100 gsm | 0.00012 | 0.21 | 0.0210 | 3.02 | 3[a] | 0.076 |
| Thin mica plastic sheet | 0.00007 | 0.212 | 0.0186 | 0.61 | 3.55[b] | 0.049 |
| Thick mica plastic sheet | 0.00019 | 0.215 | 0.0517 | 12.3 | 4.00[b] | 0.099 |
| Buffalo branding-name paper | 0.00038 | 0.209 | 0.0465 | 95.6 | 3.20[a] | 0.188 |

(a) Data from Refs. [21,22], (b) Data from Ref. [20]

From Eq. (15), we clearly can conclude that when the length of the slender column having uniform $Y$, $I$, and $\lambda$ is scaled with $L_0$ then the bending profile of all columns coincide. We need only one bending profile for all columns when the fixed end boundary is identical.

**Figure 5** shows the comparison of bending profile of mica plastic sheet and buffalo branding-name paper at the same length in the scale of characteristic length. By comparing both papers bending at three lengths: $2\tilde{L}$, $2.3\tilde{L}$, and $2.5\tilde{L}$, it is clearly observed that they show precisely identical bending.

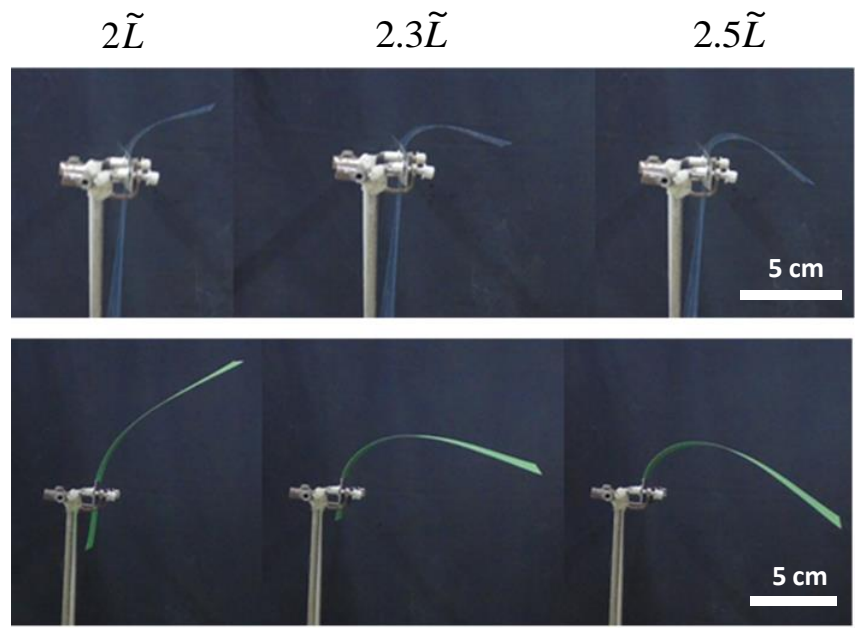

*Figure 5* Bending profile of: (upper) a mica plastic sheet and (bottom) a bufallo branding-name paper at the same length in the scale of characteristic length: (left) $2\tilde{L}$, (middle) $2.3\tilde{L}$, and (right) $2.5\tilde{L}$.

Let us look back at Eq. (13) which can be written as $L_{cr} = (N_{cr}^3 \xi YI / \lambda g)^{1/3}$. For columns having constant λ and Y, the critical length for buckling is solely controlled by area moment, and the area moment itself depends on the column geometry. If the geometry of the column can be changed even the length is kept to be constant, the phase transition can also be generated. It is likely a mechanism that controls the bending profile of tree leaves having a geometry similar to slender columns such as *banana* leaves, palm leaves, and *pandanus* leaves. Discussion of the leaf as cantilever column has been reported by Pasini and Mirjalili [24] and Ennos et al [25]. Their discussions were focused on the effect of petiole on the leaf bending, and neither related their discussion with phase transition.

The *banana* leaves (**Figure 6(a)**) start to grow in nearly cylindrical shape (rolled sheet), the cross section of which is illustrated in **Fig. 6(b)** with an effective radius *r* and the are moment approximated by $I = \pi r^4 / 4$. After a few days, the cylinder opens and the shape changes into the sheet and the area moment is around $I' = w\delta^3 / 12$ (we approximate the shape with a rectangular sheet), with w and δ are the leaf width and thickness, respectively.

The cross section area of the young banana leaf is $\pi r^2$. Since the sheet thickness is δ, the width of the sheet after opening becomes $w = \pi r^2 / \delta$. Therefore, the area moment after the leaf opens is approximately $I' = (\pi^2 / 3) I (r/w)^2$, and by using Eq. (14) we obtain

$$L_{cr} \propto \left(\frac{r}{w}\right)^{2/3} \qquad (16)$$

It is clear from Eq. (16), the leaves bend larger when the the radius to width ratio is smaller. We can observe in many images available in the internet or directly in the garden, that wider leaves bend larger than slim leaves. A Similar mechanism also happens in bending of palm frond, where the young one shapes like a vertical rod (the area moment is large). After few days, the leaves open to decrease the area moment, so that the critical length for buckling is surpassed, and the frond becomes bending.

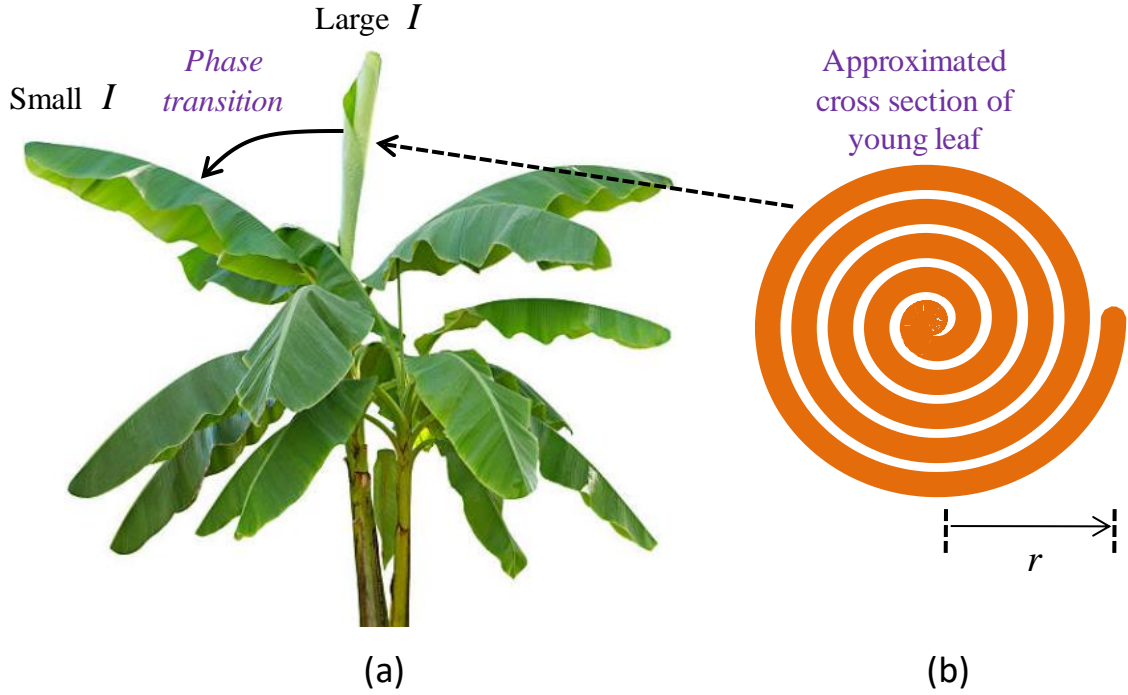

**Figure 6**. Picture of (a) banana leaves and (b) approximated cross section of young banana leaf that is assumed as rolled sheet.

For special case when the load weight is much larger than the column weight, Eq. (1) approaches the *Euler strut* as discussed by Bobnar et al [26]. However, if the load at the free end is absent, Eq. (1) estimates precisely the same critical length for buckling as have been well understood. Proves of these two statements are shown in **Appendix 2**.

## 4. Conclusion

We have demonstrated clearly that self-buckling of inextensible slender columns has a strong relationship with the second phase transition. The "critical temperature" for the phase transition is represented by the inverse of the critical height for buckling, $1/L_{cr}$, and the deviation angle made by column free end relative to vertical direction satisfies a scaling relationship $|\Delta\theta_1| \propto (1/L_{cr} - 1/L)^{0.485}$. We have also introduced a universal equation applied to all slender columns having homogeneous elastic modulus, area moment, and mass density. Using this universal equation, we re-derived the famous critical buckling height of $L_{cr} = (7.8373 YI / \rho g A)^{1/3}$ that has been derived two centuries ago using Bessel function.

Finally, we introduced a new characteristic length $L_0 = (YI/g\lambda)^{1/3}$, and if the column length is scaled with this characteristic length, the bending profile of all columns that are positioned at the same fix angle are precisely coincided. This scale unifies all kinds of inextensible and homogeneous slender columns.

**Appendix 1**

We can write Eq. (10) as following

$$\theta_{j+1} = \theta_j - \frac{a^3\lambda g}{YI}\sum_{i=1}^{j} i\cos\theta_i.$$

Or

$$\theta_{j+1} - \theta_j = -\frac{a^3\lambda g}{YI} j\cos\theta_j - \frac{a^3\lambda g}{YI}\sum_{i=1}^{j-1} i\cos\theta_i. \tag{A1}$$

Substituting Eq. (10) into (A4) we have

$$\frac{(\theta_{j+1} - \theta_j) - (\theta_j - \theta_{j-1})}{a^2} = -\frac{\lambda g}{YI}(ja)\cos\theta_j. \tag{A2}$$

If the number of segments id very large, then $a \to 0$. It causes the left hand side of Eq. (A2) to become the perfect second order derivative of θ at position *ja* from the free end, or

$$\left.\frac{d^2\theta}{ds^2}\right|_{ja} = -\frac{\lambda g}{YI}(ja)\cos\theta_j. \tag{A3}$$

Now, let us change the number of segment to become *N'* = *pN*. This changes causes the segment length to become *a'* = *a*/*p*. The index of a certain position at the rod that initially was *j* changes to *pj*. Thus the second derivate at the same position after increasing the number of segments becomes

$$\left.\frac{d^2\theta}{ds^2}\right|_{(pj)(a/p)} = -\frac{\lambda g}{YI}((pj)(a/p))\cos\theta_{pj} = -\frac{\lambda g}{YI}(ja)\cos\theta_{pj}. \tag{A4}$$

Eq. (A4) mentions that the second derivative at the same position does not change when changing the number of segment to become p fold. This concludes that the obtained profile of bending road does not change if we change the number of segment in Eq. (10).

**Appendix 2**

The discrete form of bending equation as given by Eq. (1) can be transformed into continuous form as following. By increment the index by one and assuming the load is placed at the free end, we can rewrite Eq. (1) as

$$\theta_{j+1} \approx \theta_j - \frac{a^2 W_1}{Y_j I_j}\sum_{i=1}^{j}\cos\theta_i - \frac{a^3 g}{Y_j I_j}\sum_{i=1}^{j}\sum_{k=1}^{i}\lambda_k \cos\theta_i \tag{A5}$$

In expression (A5) we have approximated $Y_i \approx Y_{i+1}$ and $I_i \approx I_{i+1}$, by choosing $a \to 0$ so that the nearest neighbor parameters have nearly the same value. Subtracting Eq. (A5) with Eq. (1) and rearranging one has

$$\frac{\theta_{j+1} - 2\theta_j + \theta_{j-1}}{a^2} \approx -\frac{W_1}{Y_j I_j}\cos\theta_j - \frac{g}{Y_j I_j}\cos\theta_j \sum_{k=1}^{i}\lambda_k a \tag{A6}$$

Eq. (A6) can be transformed into continuous form as follows

$$\frac{d^2\theta}{ds^2} \approx -\frac{W_1}{Y(s)I(s)}\cos\theta(s) - \frac{g}{Y(s)I(s)}\cos\theta(s)\int_0^s \lambda(s')ds' \tag{A7}$$

Eq. (A7) is the general expression of slender rod bending of arbitrary mass distribution, area moment, and elastic modulus.

Let us consider several special cases. First, when the load weight is much larger than the column weight so that $W_1 \gg g\int_0^L \lambda(s')ds' > g\int_0^s \lambda(s')ds'$. We obtain the approximated equation $d^2\theta/ds^2 \approx -(W_1/Y(s)I(s))\cos\theta(s)$. If $Y$ and $I$ are constant, we have $d^2\theta/ds^2 \approx -(W_1/YI)\cos\theta(s)$. The solution of this equation has been discussed in detail by Bobnar et al for *Euler strut* design, resulting in that $\theta(s)$ can be expressed by a Jacoby elliptic function [26]

If the load is absent, we can write Eq. (A7) as $d^2\theta/ds^2 \approx -(g/Y(s)I(s))\cos\theta(s)\int_0^s \lambda(s')ds'$. Furthermore, in a special case when the elastic modulus, area moment, and mass density are constant, this equation reduces to become

$$\frac{d^2\theta}{ds^2} \approx -\frac{g\lambda}{YI} s\cos\theta \ . \tag{A8}$$

We need to inspect the condition of buckling, where the column that positioned vertically starts to deviate from the vertical direction. At the vertical position, $\theta = -\pi/2$. When the column starts to exhibit buckling, the angle at the coordinate $s$ becomes $\theta(s) = -\pi/2 + \phi(s)$ where $|\phi(s)| \ll 1$ (right side of **Fig. 2**). Using the trigonometrical identity we can write $\cos\theta = \sin\phi$. Also, based on this transformation we have the following identity $d^2\theta/ds^2 = d^2\phi/ds^2$. Furthermore, since $|\phi(s)| \ll 1$ we can approximate $\sin\phi \approx \phi$ and Eq. (A8) can be approximated as

$$\frac{d^2\phi}{ds^2} + \frac{g\lambda}{YI} s\phi = 0 \tag{A9}$$

The general solution of Eq. (A9) is [27]

$$\phi(s) = C_1 Ai\left((-g\lambda/YI)^{1/3} s\right) + C_2 Bi\left((-g\lambda/YI)^{1/3} s\right) \tag{A10}$$

Since the column is positioned vertically and the angle approaches zero, we always have $\phi(0) = 0$. Therefore $0 = C_1 Ai(0) + C_2 B_i(0)$. Since $Ai(0) = 1/(3^{2/3}\Gamma(2/3))$ and $Bi(0) = 1/(3^{1/6}\Gamma(2/3))$ we then have $C_2 = -C_1/\sqrt{3}$, and the solution for $\phi$ becomes

$$\phi(s) = C_1\left(Ai\left((-g\lambda/YI)^{1/3} s\right) - \frac{1}{\sqrt{3}} Bi\left((-g\lambda/YI)^{1/3} s\right)\right) \tag{A11}$$

Then using the relationship between the Airy function and the Bessel function where $Ai(-x)$ and $Bi(-x)$ is a superposition of $J_{1/3}(2x^{3/2}/3)$ and $J_{-1/3}(2x^{3/2}/3)$ we can prove the following equation

$$\phi(s) = 2C_1 \sqrt{\frac{(g\lambda/YI)^{1/3}s}{9}} J_{1/3}\left(2\sqrt{g\lambda/YI}\, s^{3/2}/3\right) \quad \text{(A12)}.$$

The second boundary condition is curvature at the free end is zero or $d\phi/ds]_{s=L} = 0$, resulting $d/ds\left(\sqrt{s} J_{1/3}\left(2\sqrt{g\lambda/YI}\, s^{3/2}/3\right)\right) = 0$. Let us temporarily suppose $z = (2/3)\sqrt{g\lambda/YI}\, s^{3/2}$ so that the above identity can be written as $d/ds\left(z^{1/3} J_{1/3}(z)\right) = 0$ or $d/dz\left(z^{1/3} J_{1/3}(z)\right) dz/ds = 0$ to result $d/dz\left(z^{1/3} J_{1/3}(z)\right) = 0$. We then use the following identity $(1/z\, d/dz)^m \left(z^\alpha J_\alpha(z)\right) = z^{\alpha-m} J_{\alpha-m}(z)$, where in our condition $m = 1$ and $\alpha = 1/3$ so that $(1/z\, d/dz)\left(z^{1/3} J_{1/3}(z)\right) = z^{-2/3} J_{-2/3}(z)$ or $d/dz\left(z^{1/3} J_{1/3}(z)\right) = z^{1/3} J_{-2/3}(z)$. Thus to ensure zero, z must be the root of $J_{-2/3}$ or $J_{-2/3}(z_0) = 0$. This results the expected critical height for buckling, $L_{cr} = (3/2)^{2/3} z_0^{2/3} (YI/\lambda g)^{1/3}$ $L_{cr} = (3/2)^{2/3} z_0^{2/3} (YI/\lambda g)^{1/3}$, similar to that previously obtained.

**Acknowledgement**

PMDSU Fellowships from the Ministry of Research and Higher Education Republic of Indonesia for DMO, NA, and FDU are highly acknowledged. This work was also supported by Research Grant from Bandung Institute of Technology 2019 for M.A.

**Disclosure statement**

**References**

[1] Deegan RD, Bakajin O, Dupont TF, Huber G, Nagel SR, Witten TA. Capillary flow as the cause of ring stains from dried liquid drops. Nature 1997;389:827–829.

[2] Lee A, Brun P-T, Marthelot J, Balestra G, Gallaire F, Reis PM. Fabrication of slender elastic shells by the coating of curved surfaces. Nature Commun. 2016;7:11155.

[3] Jawed MK, Dieleman P, Audoly B, Reis PM. Untangling the mechanics and topology in the frictional response of long overhand elastic knots. Phys Rev Lett. 2015;115:118302.

[4] Miller JT, Lazarus A, Audoly B, Reis PM. Shapes of a suspended curly hair. Phys Rev Lett. 2014;112:068103.

[5] Lazarus A, Miller JT, Reis PM. Continuation of equilibria and stability of slender elastic rods using an asymptotic numerical method. J Mech Phys Solids. 2013;61:1712(2013).

[6] Akono A-T, Reis PM, Ulm F-J. Scratching as a fracture process: from butter to steel. Phys Rev Lett. 2011;106:204302.

[7] Vandeparre H, Pineirua M, Brau F, Roman B, Bico J, Gay C, Bao W, Lau CN, Reis PM, Damman P. Wrinkling hierarchy in constrained thin sheets from suspended graphene to curtains. Phys Rev Lett. 2011;106:224301.

[8] Reis PM, Hure J, Jung S, Bush JWM, Clanet C. Grabbing water. Soft Matter. 2010;6: 5705.

[9] Bîrsan M, Altenbach H, Sadowski T, Eremeyev VA, Pietras D. Deformation analysis of functionally graded beams by the direct approach. Composites: Part B. 2012;43:1315–1328.

[10] Bîrsan M, Sadowski T, Marsavina L, Linul E, Pietras D. Mechanical behavior of sandwich composite beams made of foams and functionally graded materials. Int J Solids Struct. 2013;50:519–530.

[11] Linul E, Marşavina L. Assesment of sandwich beams with rigid polyurethane foam core using failure-mode maps, Proc Romanian Acad Series A. 2015;16:522–530.

[12] Linul E, Valean C, Linul P-A. Compressive behavior of aluminum microfibers reinforced semi-rigid polyurethane foams` Polymers. 2018;10:1298.

[13] Linul E, Marsavina L, Voiconi T, Sadowski T. Study of factors influencing the mechanical properties of polyurethane foams under dynamic compression, J Phys: Conf Ser. 2013;451:012002.

[14] Berto F, Razavi SMJ, Ayatollahi MR. Fatigue assessment of steel rollers using an energy based criterion. Procedia Struct Integrity. 2017;3:93-101.

[15] Berto F, Ayatollahi MR, Campagnolo A. Fracture tests under mixed mode I+III loading: An assessment based on the local energy. Int J Damage Mech. 2017;26:881-894.

[16] Zeyad AD. SViscoelastoplastic bodies under cyclic loading in thermal-radiation fields. J Taibah Univ Sci. 2017;11:605–612.

[17] Greenhill A. Determinationm of the greatest height consistent with stability that a vertical pole or must can be made, and of the greatest height to which a tree of given proportion can grow. Proc Camb Philol Soc.1881; 4: 65–73.

[18] Rotter JM, Sadowski AJ, Chen L. Nonlinear stability of thin elastic cylinders of different length under global bending. Int J Solids Struct. 2014;51:2826–2839.

[19] Rotter JM, Sadowski AJ. Full plastic resistance of tubes under bending and axial force: exact treatment and approximations. Structures. 2017;10:30–38.

[20] Amalia N, Yuliza E, Margaretta DO, Utami FD, Surtiyeni N, Viridi S, Abdullah M. A novel method for characterizing temperature-dependent elastic modulus and glass transition temperature by processing the images of bending cantilever slender beams at different temperatures. AIP Advances. 2018; 8: 115201.

[21] Okomori K, Enomae T, Onabe F. Evaluation and control of coated paper stiffness. in Proc Tappi Adv Coat Fundam Symp. (1999) 121–132.

[22] Ek G, Gellerstedt G, Henriksson G. Paper Products Physics and Technology. Berlin:Walter de Gruyter;2009.

[23] Landau LD, Lifshitz EM. Statistical Physics. Volume 5;3rd edition. Oxford: Butterworth-Heiemann;1980.

[24] Pasini D, Mirjalili V. The optimized shape of a leaf petiole. Comparing Design Nat Sci Eng. 2006; 87: 35–45

[25] Ennos AR, Spatz H-Ch, Speck T. The functional morphology of the petioles of the banana, *Musa textilis*. J Experiment Biol. 2000; 51, 2085-2093.

[26]   Bobnar J, Susman K. et al. Euler strut: A mechanical analogy for dynamics in the vicinity of a critical point. Eur J Phys. 2011; **32:**1007.

[27]   Wolfram Mathematica;2018.